\documentclass{PoS}
\usepackage{cite}
\usepackage{wrapfig,rotating}

\title{Parton densities from DIS and hadron colliders to LHC: WG summary}

\ShortTitle{Parton densities from DIS and hadron colliders to LHC: WG summary}

\author{ {\speaker {Sergey Alekhin$^{1}$} },  Dimitri Colferai$^{2}$, 
    Joey Huston$^{3}$ and {\speaker {Ringail\. e Pla\v cakyt\. e$^{4}$} }%
\\ \\
1 - Deutsches Elektronen-Synchrotron DESY, \\
Platanenallee 6, D-15738 Zeuthen, Germany  \& \\
Institute for High Energy Physics IHEP, \\ 
Pobeda 1, 142281 Protvino, Russia. \\ 
E-mail: \email{sergey.alekhin@ihep.ru} \\ \\
2 - University of Florence, \\
Department of Physics, \\
Via G. Sansone 1, I-50019 Sesto Fiorentino (Firenze), Italy. \\
E-mail: \email{colferai@fi.infn.it} \\ \\
3 - Michigan State University, \\
Department of Physics and Astronomy, \\
3218 Biomedical Physical Science Building, \\
East Lansing, MI 48824, USA. \\
E-mail: \email{huston@pa.msu.edu} \\ \\
4 - Deutsches Elektronen Synchrotron DESY, \\    
Notkestrasse 85, D-22607 Hamburg, Germany. \\
E-mail: \email{ringaile@mail.desy.de}   \\
}

\abstract{Recent experimental and theoretical results presented in the
  working group  
 "Parton densities from DIS and hadron colliders to LHC" at the DIS2010 
  workshop are summarized in this contribution.}
\FullConference{XVIII International Workshop on Deep-Inelastic Scattering and Related Subjects\\
                 April 19 -23, 2010\\
                 Convitto della Calza, Firenze, Italy}

\begin{document}

Parton distribution functions (PDFs) are determined from global analyses of a wide range 
of data resulting from a variety of hard-scattering processes; these processes include deep inelastic
scattering (DIS) of leptons off of a nucleon, lepton pair production in hadron-hadron collisions (the Drell-Yan process), and jet production in proton-antiproton 
and lepton-proton collisions (see Table~\ref{tab:pdfs}).  
The data used in these global analyses are continually updated, taking advantage both of improvements to the data sets and of progress in the theory.
At this workshop, new experimental results from HERA, Fermilab, and JLAB
experiments were presented and analyzed in the context of PDF determination~\cite{sumtalk}.
\\
\begin{table*}[th]
\newcommand{\m}{\hphantom{$-$}}
\newcommand{\cc}[1]{\multicolumn{1}{c}{#1}}
\renewcommand{\tabcolsep}{1.5pc} 
\begin{tabular}{@{}lllll}
\hline
Name &  Data used & QCD order & Scheme & Reference \\  \hline

ABKM &  DIS+DY & NLO/NNLO & FFN & \cite{ABKM} \\
CTEQ &  DIS+DY+jets & NLO & GMVFN & \cite{CTEQ} \\
HERAPDF &  DIS & NLO & GMVFN & \cite{HERACOMB} \\
JR &  DIS+DY & NLO/NNLO & FFN & \cite{JR} \\
MSTW &  DIS+DY+jets & NLO/NNLO & GMVFN & \cite{MSTW} \\
NNPDF &  DIS+DY+jets & NLO & ZMVFN & \cite{NNPDF} \\
\hline
\end{tabular}\\[2pt]
\caption{Recently published nucleon PDF sets, with a brief description of the 
data used in the fit, the theoretical accuracy and the factorization 
scheme employed to model the heavy-quark DIS contribution. 
The table is taken from Ref.~\cite{LL10}.
}
\label{tab:pdfs}
\end{table*}

The large sample of data accumulated in the fifteen years of successful HERA
operation provides a comprehensive view of the proton, particularly at  
small values of the Bjorken variable~$x$. 
The latest HERA results presented in this DIS2010 workshop include 
(1) the combination of  DIS cross sections measured 
at various proton-beam energies, (2) the extraction of the longitudinal structure 
function $F_L$, and (3) updated neutral current (NC) and charged current (CC) measurements.
Recent HERA data measurements have resulted in improved determination
of both PDFs and of the weak coupling constants.

\begin{figure}[!ht]
 \begin{minipage}[b]{0.5\linewidth} 
   \centering
  \includegraphics[width=7.4cm]{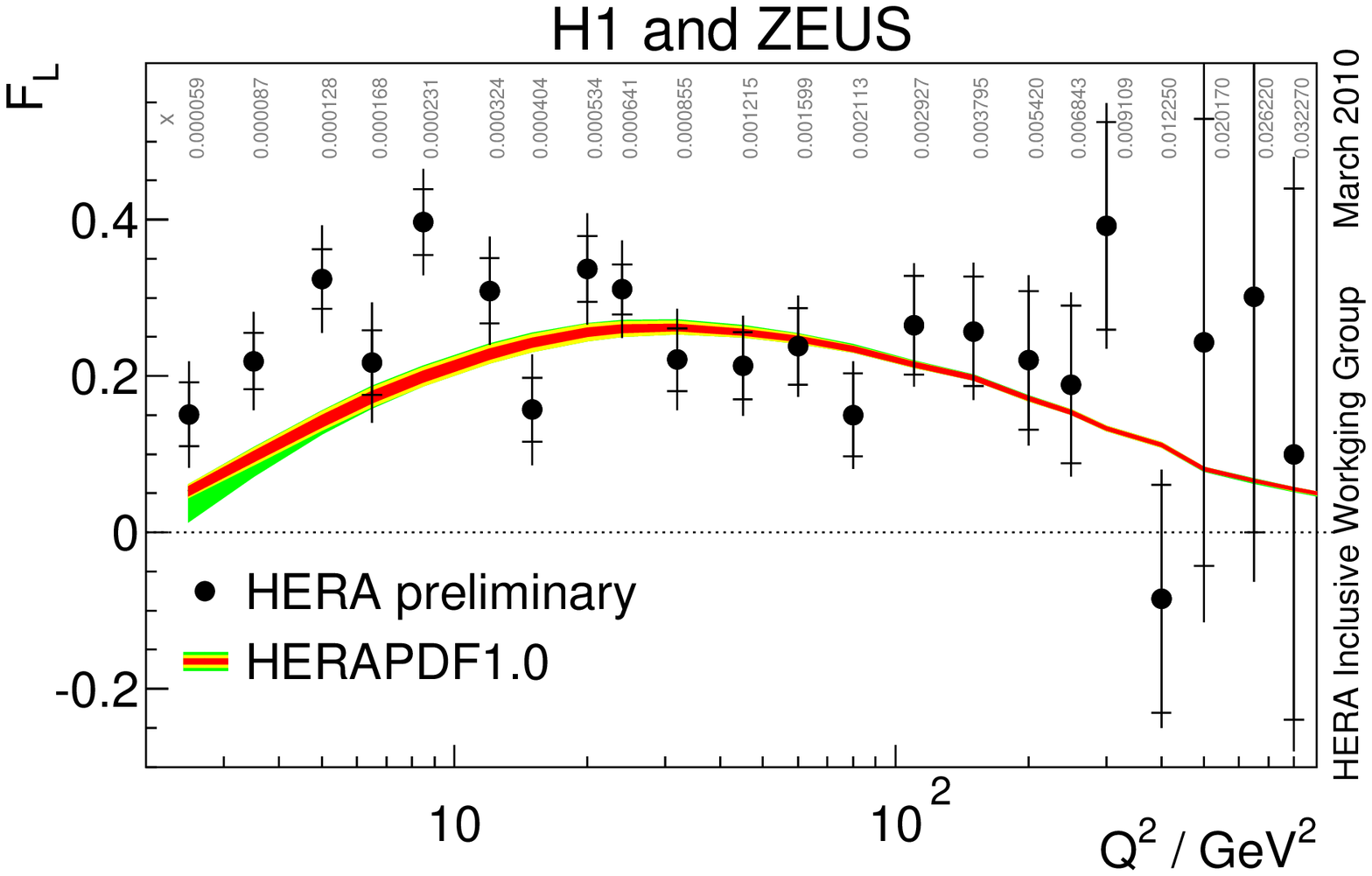}
 \end{minipage}
 \begin{minipage}[b]{0.5\linewidth} 

   \centering
  \includegraphics[width=7.4cm]{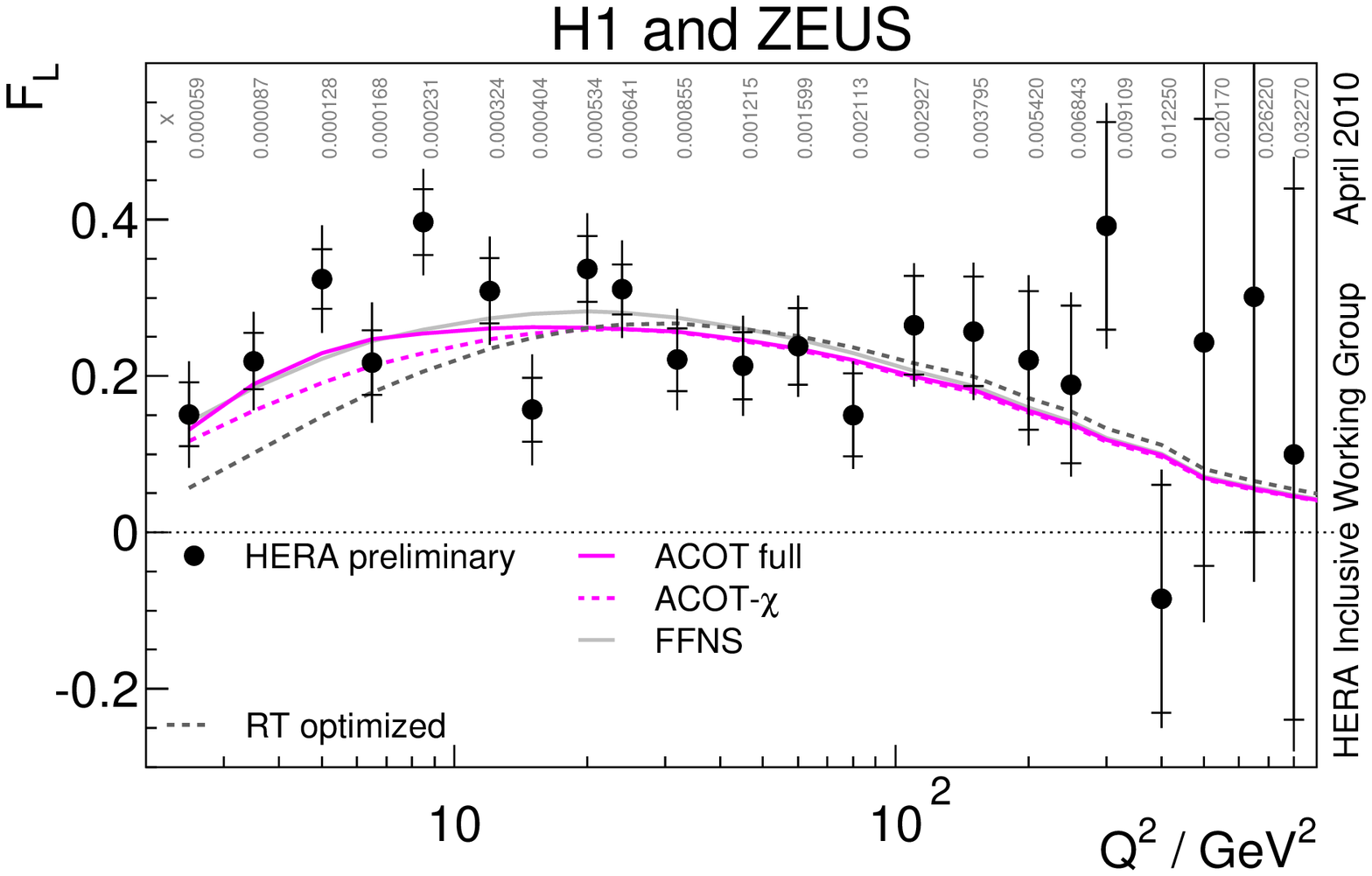}
 \end{minipage}
  \caption{The combined H1 and ZEUS measurement of
               the structure function $F_L$, averaged in x at a given value of $Q^2$,
               compared to a QCD prediction based on the PDFs of Ref.~\cite{HERACOMB} ({\bf left}) 
               and different variants of the fits of Ref.~\cite{radescu1} {\bf (right)}. 
}
  \label{combFl}
\end{figure}

A new combined data set on  inclusive $e^{+}p$ DIS cross sections 
measured at different proton-beam energies
($E_p = 920$ GeV, $E_p = 460$ GeV and $E_p$= $575$ GeV) was recently produced by the H1 and 
ZEUS Collaborations~\cite{loweprelim}. 
The variation of the centre-of-mass energy~$\sqrt s$ 
allows for the discrimination of the structure functions $F_L$ and $F_2$.
Since DIS cross sections are
proportional to $F_2 -y^2 F_L/ (1+(1-y)^2)$, 
$F_L$ and $F_2$ can be extracted from  measurements at two or more different 
values of inelasticity $y=Q^2/xs$.
The combination of the H1 and ZEUS data includes the cross-calibration
of two experiments, and results in an improved precision for the data. 
The proton structure function $F_L$ is extracted from the combined cross sections 
in the range of $2.5 \le Q^2 \le 800$ GeV$^2$ (see Figure~\ref{combFl}).

In addition to the combined HERA data of Ref.~\cite{loweprelim}, a new extended 
NC measurement of the $e^+p$ inclusive cross sections at different
proton-beam energies has been obtained by the ZEUS Collaboration~\cite{loweprelim}
%
%
For this analysis, satellite vertex events were used to access $Q^2$ values down
to 5~GeV$^2$ for the reduced proton beam energies.
This measurement should further improve constraints on $F_L$ provided by HERA. 

The combined data of Ref.~\cite{loweprelim} were employed in a QCD fit
taking into account  QCD corrections up to  NNLO~\cite{radescu1}.
Two variants of the fit with two different cuts on $Q^2$ were tried. In this way,
the sensitivity of the PDFs to the low $Q^2$ portion of the data was determined.
The fit is sensitive to the treatment of the heavy quark contribution
to the inclusive DIS cross sections.
In particular, the predictions based on the optimized Thorne prescription of Ref.~\cite{Thorne} 
undershoot the HERA $F_L$ measurement (see Figure~\ref{combFl}). 
The ACOT prescription~\cite{ACOTfull,ACOTchi} and the (3-flavor)
fixed-flavor number (FFN) scheme
are in better agreement with the data.
 
New NC and CC DIS cross section measurements at high $Q^2$ were
obtained by the H1 and ZEUS Collaborations using the HERA-II sample obtained with 
longitudinally polarised leptons.
These data provide additional constraints on PDFs through the NC polarisation 
asymmetries and test the chiral structure of the weak interactions.  

The ZEUS Collaboration has measured the CC cross sections based on a sample with an integrated 
luminosity of 132 $pb^{-1}$ and different polarizations of the positron beam~\cite{zeuscc}.
The H1 Collaboration has determined the CC cross sections 
using the complete HERA-II data sample with polarized $e^{+}$ and $e^{-}$ beams~\cite{h1cc}. 
In order to achieve better precision in the CC data,
the HERA-II CC polarised measurements were combined with the unpolarised HERA-I data. 
The combined sample corresponds to the total luminosity of 280.8 $pb^{-1}$ and 
165.5 $pb^{-1}$ for $e^+p$ and $e^-p$ scattering, respectively. 

The NC double-differential cross sections
for $e^{\pm}p$ scattering with longitudinally polarised lepton beams
were measured by the H1 Collaboration~\cite{h1nc}. 
The charge dependent polarisation asymmetry in the NC cross sections 
is sensitive to the quark vector and axial couplings to the Z boson. 
Similar to the case of CC interactions, the NC cross sections
were combined with the earlier H1 measurements.
The structure functions $x \tilde F_3$ and $xF_3^{\gamma Z}$ were extracted
from the combined unpolarised cross section measurements.

A new method to measure the NC cross sections up to values of $x$ close to 1
was employed by the ZEUS Collaboration~\cite{zeusnc}.
In this region, PDFs are poorly known and the data of Ref.~\cite{zeusnc} provide valuable 
additional constraints for the global PDF fits. 
The NC cross sections were extracted at $Q^2 \ge 575$ GeV$^2$ using 
the $e^-p$ collision data sample with an integrated luminosity of 187 $pb^{-1}$.
This is a factor of ten larger than the one used for the  earlier ZEUS measurement~\cite{zeusnchera1}.
The systematic uncertainties are also reduced as 
compared to Ref.~\cite{zeusnchera1} 
due to improved kinematic reconstruction methods.
The data of Ref.~\cite{zeusnc} were found to be in good agreement with the 
predictions based on the CTEQ6D PDFs.

\begin{figure}[!ht]
 \hspace{0.7cm} 
 \begin{minipage}[b]{0.5\linewidth} 
   \centering
   \includegraphics[width=7cm]{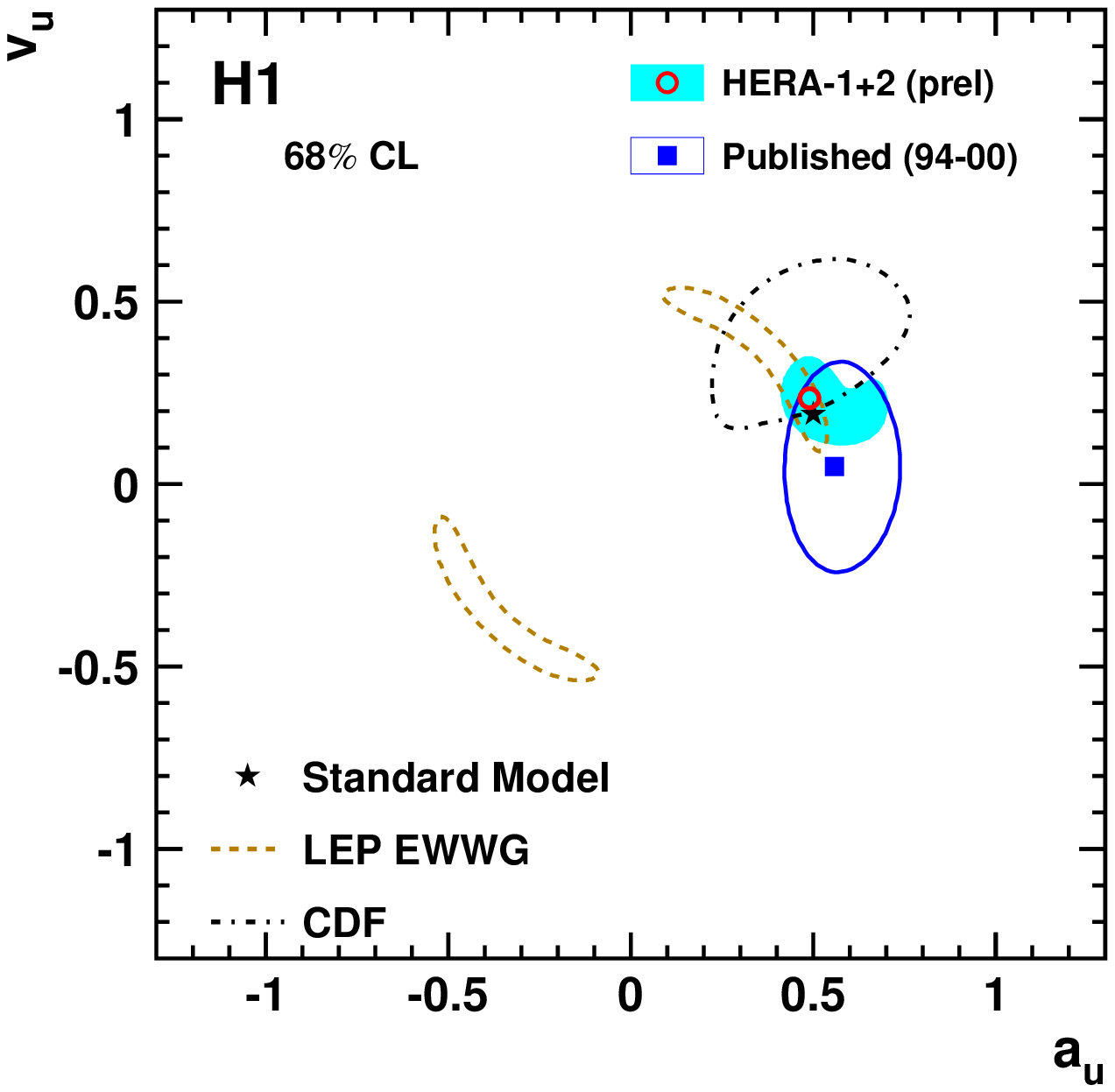}
 \end{minipage}
 \hspace{-1.4cm} 
 \begin{minipage}[b]{0.5\linewidth}
   \centering
   \includegraphics[width=7cm]{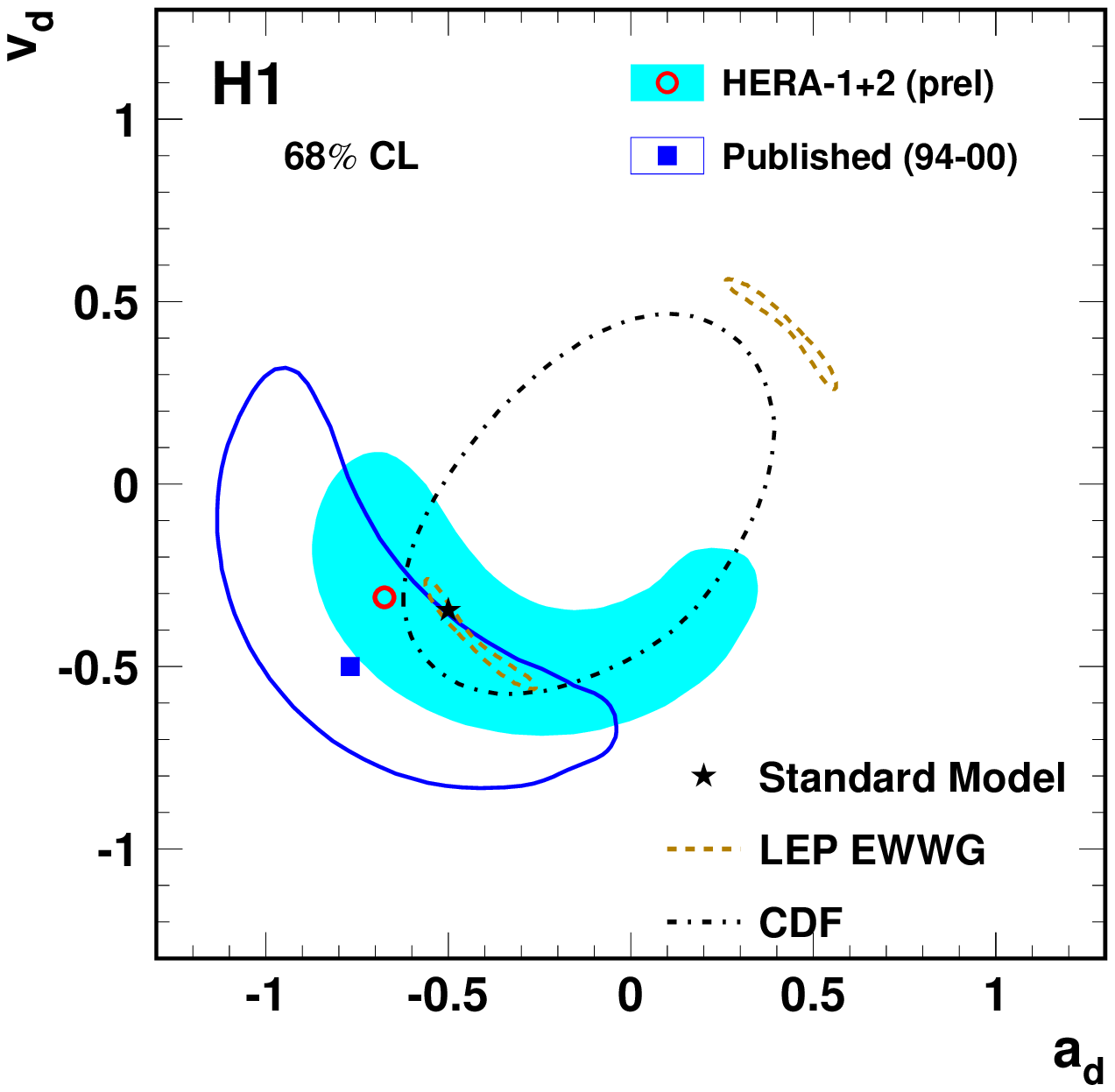}
 \end{minipage}
 \caption{
         The 68$\%$ confidence level for the determination of the weak couplings of 
         up-({\bf left}) and down-type({\bf right})
         quarks to the $Z$ boson determined in a combined electroweak-QCD 
         analysis using the full HERA H1 data~\cite{h1nc}; the results are compared with 
         the corresponding results published previously using the HERA-I data 
         alone~\cite{ewqcdhera1}; couplings determined by the LEP~\cite{lepcouplings}
         and TEVATRON~\cite{cdfcouplings} are given for the comparison.}
 \label{qcdewfits}
\end{figure}

A new combined electroweak and QCD analysis was performed by the H1 Collaboration
using the full low and high $Q^2$ HERA data~\cite{zhang}.
Weak vector and axial couplings of light up- and down-type quarks to the $Z$ boson ($v_q, a_q$) 
were extracted from the combined fit simultaneously with the PDFs.
Due to the additional sensitivity of the polarised NC measurement to the 
quark vector coupling, the accuracy of the vector coupling $v_q$ was improved with 
respect to earlier results based
on the unpolarised HERA data only~\cite{ewqcdhera1}. 
The constraint obtained on the up-type quark coupling $v_{u}$ from HERA is better than those obtained from 
LEP~\cite{lepcouplings} and from the TEVATRON~\cite{cdfcouplings} (see Figure~\ref{qcdewfits}). 
Moreover, the HERA data are also sensitive to the sign of weak couplings.

The H1 Collaboration presented proton structure function measurements at low and 
medium $Q^2$ determined from the first period of HERA operation. 
The inclusive double differential cross sections 
for $e^{+}p$ scattering were obtained at 12~$\le~Q^2~\le~150$~GeV$^2$ \
for an integrated luminosity of 22~pb$^{-1}$~\cite{klein}.
The data were collected at the beam energies of E$_e$~=~27.6~GeV and E$_p$~=~920 GeV 
and were combined with earlier data obtained with E$_p$ = 820 GeV.
The typical precision of the combined measurements is $1.3-2\%$.
This is a record level of  accuracy for  DIS measurements.
The NLO QCD fit to the H1 data based on the new measurements provides an improved  
determination of the gluon and quark densities in the proton, particularly at 
small~x~\cite{medQ2paper}.
The analysis of Ref.~\cite{medQ2paper} includes the study of the experimental, 
model and parameterisation uncertainties in PDFs.

The inclusive NC double differential cross sections of
$e^{+}p$ scattering at HERA were obtained by the H1 Collaboration at 
small $x$ and low $Q^2$ ($0.2 \le Q^2 \le 12 $~GeV$^2$)~\cite{glazov,lowQ2paper}.
These data were collected in two dedicated periods with nominal and 
shifted interaction vertex, where a shifted vertex  provided better 
acceptance at low $Q^2$.
The two samples overlap at $0.5~\le~Q^{2}~\le~3.5$ GeV$^2$, 
and the proton structure functions $F_2$ and $F_L$ were extracted from the combination 
of these measurements.
The data at $Q^2 \lesssim 2$ GeV$^2$ correspond to the transition between DIS 
and photoproduction regimes. With the improved data accuracy, one can discriminate 
between different theoretical
approaches used to model $F_{2}$ and $F_{L}$ in this region~\cite{lowQ2paper}.
 
The inclusive $e^{\pm}p$ NC and CC cross sections measured by the H1 and ZEUS Collaborations 
in the first period of HERA operation were combined into a common data set~\cite{habib}. 
The combined data cover the range of $6 \cdot 10^{-7} < x < 0.65$,
$0.045 \le Q^2 \le 30000$ GeV$^2$ for the NC and  $0.013 < x < 0.4$, 
$300 \le Q^2 \le 30000$ GeV$^2$ for the CC scattering, respectively.
The H1 and ZEUS input data used in the combination were found to be
consistent with each other with $\chi^2/DOF = 636.5/656$.
The total uncertainty reached with this combination is 1$\%$ for  NC interactions in the most accurately 
measured region ($20 < Q^2 <100$~GeV$^2$)~\cite{HERACOMB}. 
A NLO QCD analysis of the combined $e^{\pm}p$ scattering cross-section data was performed
and a new set of parton distribution functions, HERAPDF 1.0, was obtained from the analysis~\cite{HERACOMB}.
The experimental uncertainties in the PDFs were reduced
with respect to the earlier H1 and ZEUS PDF sets, due to the improved accuracy in the combined data.
The theoretical uncertainties obtained by varying the input assumptions of the fit, 
e.g. the charm quark mass, were also studied. 
\\
\begin{wrapfigure}{r}{0.5\columnwidth}
\centerline{\includegraphics[width=0.4\textwidth,angle=-90]
{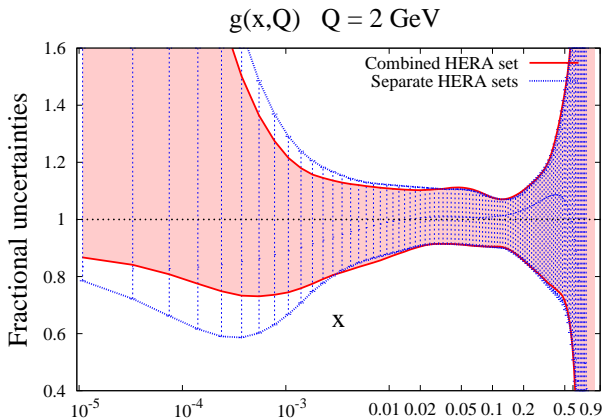}}
\caption{Impact of the combined inclusive HERA data on the 
fractional gluon uncertainty. Figure taken from Ref.~\cite{CTEQ10}.} 
\vspace{-0.5cm}
\label{fig:cteq10}

\end{wrapfigure}

Several groups~\cite{MSTW-DIS,CTEQ10,NNPDF,ABKM-DIS}
have recently replaced the inclusive H1 and ZEUS DIS data
of Refs.~\cite{H1ZEUS} used in their PDF fits with 
the combined HERA data of Ref.~\cite{HERACOMB}. 
The new HERA data are in reasonable agreement with the other data sets
used in the fits. Moreover, this replacement leads to an improvement 
in the small-$x$ gluon and sea uncertainties, 
typically by the order of 20-30\% (see Figure~\ref{fig:cteq10}).
At scales of  $O({\rm GeV}^2)$, the accuracy in the HERA data 
is better than 2\%. Due to such precision, NNLO QCD
corrections are needed for the most accurate interpretation of these data.
The NNLO (3-loop) corrections to the PDF evolution 
have been calculated in Ref.~\cite{Moch:2004pa}. In addition, 
the 3-loop corrections to the massless 
DIS coefficient functions are also 
known~\cite{Vermaseren:2005qc}. 
In the factorization scheme with 3 light flavors in the 
initial state (the FFN scheme), 
the corrections to the heavy-quark contribution to DIS  
are only known up to  NLO~\cite{Laenen:1992zk}. This makes an NNLO analysis 
of the DIS data somewhat inconsistent. The higher-order QCD 
corrections are partially taken into account through the massless 
evolution of the heavy quarks, which is employed in 
the 
zero-mass variable-flavor-number (ZMVFN) scheme. However, the ZMVFN scheme is applicable only at  
asymptotically large transfers $Q\gg m_{\rm h}$, 
where $m_{\rm h}$ denotes the heavy quark mass. For realistic kinematics,
the exact results are generally overestimated, due to the missed power  
corrections to the heavy-quark production 
coefficient functions. 
This shortcoming is overcome in the
general-mass variable-flavor-number (GMVFN) extensions of the ZMVFN scheme
by a special modeling of the Wilson coefficients at low $Q$. In this way, 
the GMVFN scheme prescriptions of ACOT, Roberts-Thorne, and Thorne 
were obtained (cf. review of Ref.~\cite{Jung:2009eq} 
and Refs.~\cite{Nadolsky:2009ge,Thorne} for the recent update of the  
ACOT and Thorne's prescriptions).
\begin{wrapfigure}{r}{0.5\columnwidth}
\hspace{-0.5cm}
\includegraphics[width=19pc,height=17pc]{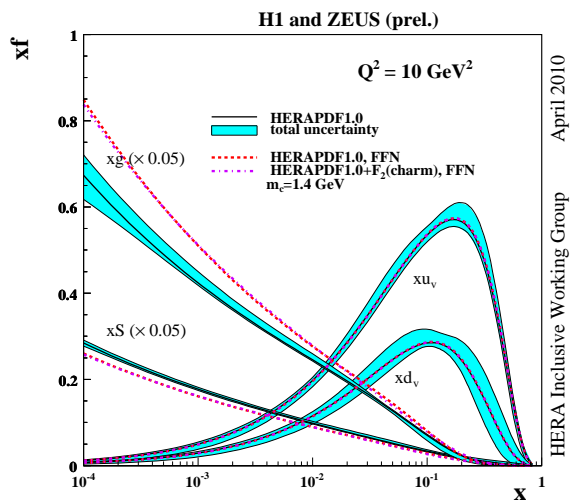}
\caption{The PDFs obtained in Ref.~\cite{HERACOMB}
for the GMVFN scheme of Ref.~\cite{Thorne} 
and in the variant of this fit 
using the FFN scheme. The figure is taken from Ref.~\cite{mandy}.
}
\label{fig:mandy}
\end{wrapfigure}

The GMVFN schemes 
facilitate combination of the DIS and the hadron collider data in the global 
PDF fits. However, the GMVFN PDFs are sensitive to the specific choice of 
the prescription. This results to an additional uncertainty in the hard
cross section predictions based on these PDFs~\cite{Tung:2006tb}. 
The FFN scheme and the different prescriptions of 
GMVFN schemes provide an equally good description of  
the combined HERA data on the inclusive and semi-inclusive 
structure functions $F_2^c$~\cite{mandy}.
Meanwhile, the FFN small-$x$ gluon distribution 
obtained in the fit of Ref.~\cite{mandy}
is substantially higher than gluon distribution determined in the GMVFN variants 
of the fit (see Figure~\ref{fig:mandy}), and is clearly 
positive at small $x$.
In part, this is explained by the definition of the GMVFN PDFs; the remaining 
discrepancy can be considered as the PDF uncertainty due to the scheme choice.
It is worth noting that the PDFs obtained in the FFN fits
correspond to the $\overline{\rm{MS}}$ scheme. This is also the case for the 
GMVFN prescription suggested by Buza-Matiounine-Smith-van Neerven
(BMSN)~\cite{BMSN} and for the FONLL prescription of Ref.~\cite{FONLL}, 
which contains the main BMSN features. This is not generally guaranteed 
for PDFs obtained within the GMVFN fit formalism~\cite{chuvakin}. 
However, the higher-order corrections to  heavy-quark DIS production  
suppress the difference between the FFN and ZMVFN schemes
at large $Q$ (see Ref.~\cite{GLUCK}). 
For the NLO case, this difference can barely be resolved by the existing 
data~\cite{ABKM} and must be even smaller taking into account the 
NNLO corrections to the matching conditions
for the 4- and 5-flavor distributions~\cite{BBK}.
The interpretation of the heavy-quark electro-production and inclusive DIS data
also depends on the heavy-quark masses values and their definition. 
The HERA data alone are not sensitive to the heavy-quark masses since the 
effect of the mass variation is compensated by changes in the 
PDFs~\cite{radescu1}. The heavy quark masses can be more precisely determined in a global fit including data from other processes. Such a determination was carried out by MSTW and a set of PDFs corresponding to different values of the heavy quark masses is provided in Ref.~\cite{MSTWHQ}. 

Hadron collider data on $W/Z$ and jet production 
provide more information for PDF determination to one 
available in the DIS data alone. They provide 
constraints 
on PDFs at large factorization scales and at large parton $x$ values and help to disentangle
the distributions of different PDF species. Moreover,  $W/Z$
production is considered as one of the primary LHC standard candle processes~\cite{dittmar}, due to the large cross section and small experimental and theoretical uncertainties. 
The $W/Z$ distributions have been calculated up to NNLO 
accuracy~\cite{DYDIST}. The higher-order corrections 
of Refs.~\cite{DYDIST} are  usually 
employed in the global PDF fits in the form of fixed 
$K$-factors~\cite{MSTW,CTEQ} applied to each data point, and updated periodically through the fitting process. 
In the NNPDF fit~\cite{NNPDF}, the higher order corrections are
applied in each fitting iteration with the use of FastKernel tool.
A similar technique was employed for the first time in the FastNLO
code for the fast calculation of the jet production cross 
sections~\cite{Kluge:2006xs} and has been developed recently
within the APPLGRID project for the NLO QCD 
analysis of the LHC data on the jet and $W/Z$ production~\cite{Carli:2010rw}.

\begin{wrapfigure}{r}{0.5\columnwidth}
\hspace{-0.5cm}
\includegraphics[width=18.5pc,height=15pc]{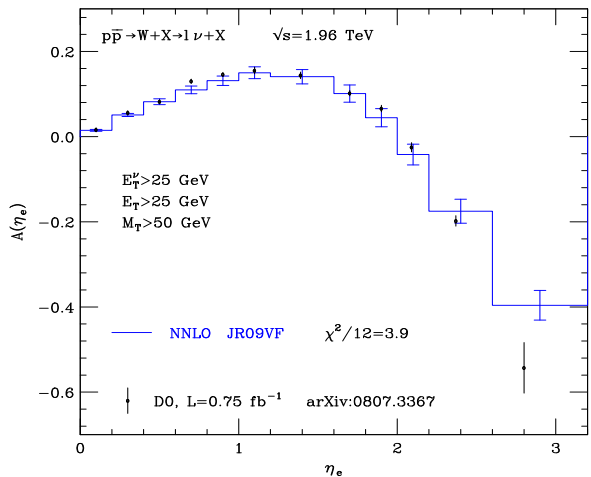}
\caption{The Run II electron charge asymmetry obtained by the D0
collaboration compared to the NNLO predictions based on the JR
PDFs of Ref.~\cite{JR}. Figure taken from Ref.~\cite{grazzini}.
}
\label{fig:wasym}
\end{wrapfigure}

Recently, precise Run II $W$ lepton asymmetry data have become available~\cite{RUNII}. The CTEQ and MSTW 
collaborations found that the 
lepton asymmetry data disagree with other data sets used in the global fitting, particularly those from the NMC~\cite{nmc} and BCDMS~\cite{bcdms1},\cite{bcdms2} experiments. The disagreement leads to a non-negligible increase in global $\chi^2$ for the global fits, and a poor agreement of the resultant predictions with the lepton asymmetry data. As a result, both CTEQ and MSTW have left the Run II lepton asymmetry data out of their latest global fits. The CTEQ and MSTW predictions, as well as those from ABKM, overshoot the data in general.  CTEQ has a variant fit (CT10W)~\cite{CTEQ10} in which the asymmetry data is included in the fit, but given a large weight. This is the only manner in which a reasonable $\chi^2$ can be obtained for the asymmetry data sets. The Run II lepton asymmetry analyses have been carried out by both CDF (for electron) and D0 (for both electron and muon). The CDF electron asymmetry data agree with the results from the D0 electron analysis. To make matters more complex, there is some tension between the D0 electron and muon asymmetry data sets. Thus, there is no final conclusion yet as to the impact of the Run II lepton asymmetry data. 
The best agreement with the data 
was obtained for the case of the JR PDFs (see Figure~\ref{fig:wasym}).  

The Tevatron charge asymmetry 
data are sensitive to the $d$-quark distribution, and to the $d/u$ quark ratio. Therefore,  the 
agreement might be improved due to modification of 
the correction for nuclear effects in deuterium, which affects the 
$d$-quark distribution extracted from the deuteron fixed target DIS data that have the conflict with the Run II lepton asymmetry data. 
A model-independent form of deuteron correction was attempted in the 
MSTW fit of Ref.~\cite{MSTW-DIS}. While this correction somewhat improves the description 
of the Tevatron charge asymmetry data, the shape of the deuteron 
correction preferred by this fit cannot be justified by a reasonable 
nuclear model.

In Ref.~\cite{Schienbein:2007fs}, nuclear corrections were also 
fitted to the charged-lepton and neutrino DIS data in the spirit of 
the nuclear PDF concept~\cite{Eskola:2009uj,Hirai:2007sx,deFlorian:2003qf}.
In this manner, different PDF shapes
were found for  neutral-current and  charged-current DIS off of an
iron target. Therefore the application of the 
resulting nuclear PDFs to collider predictions is
somewhat problematic. Meanwhile, 
the observation of Ref.~\cite{Schienbein:2007fs}
is based on the analysis of data from one 
experiment only; it therefore  requires independent confirmation. 
In the model of Ref.\cite{Kulagin:2004ie}, the various nuclear effects  
are considered separately, in contrast to the nuclear PDF approach.
This model describes a wide set of nuclear DIS data, including 
the recent JLAB data for helium-3~\cite{JLAB}. Thus, it can be conjectured
that this approach can be also 
reliably extrapolated to the case of deuteron targets for the 
benefit of interpretation of the hadron collider lepton asymmetry data. 

\begin{wrapfigure}{r}{0.5\columnwidth}
\hspace{-0.7cm}
\includegraphics[width=21pc]{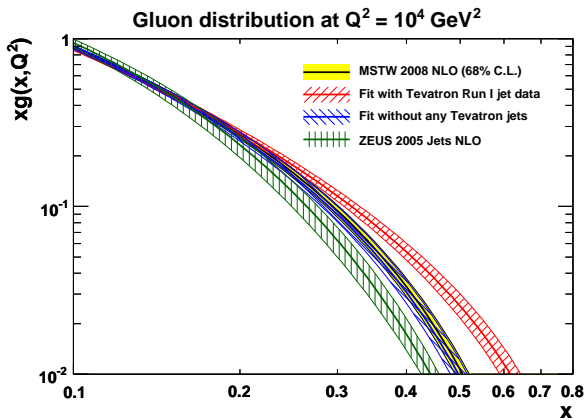}
\caption{The large-$x$ gluon distribution obtained in the different variants
of the NLO MSTW fit. Figure taken from Ref.~\cite{MSTW}. 
}
\label{fig:gluon}
\end{wrapfigure}

The production of jets at high transverse momentum in hadron collisions is sensitive to the 
large-$x$ gluon distribution at both the Tevatron and LHC colliders. The NNLO corrections to the hadronic jet 
production have not yet been calculated;  therefore the jet data can 
be consistently used only in the NLO version of the PDF fit. 
For some time, the Run I jet Tevatron data have
provided the main constraints on the high $x$ gluon distribution~\cite{MRST,CTEQ4}. 
The Run I data, and in particular the Run I data from D0, prefer a higher large-$x$ gluon distribution as compared 
to the gluon determined from fits to DIS data alone. The Run II inclusive jet 
data, especially those from D0, are relatively lower at high 
jet transverse momentum $p_T$ and jet rapidity $Y$, and the resultant global fits using this data alone have a lower gluon distribution at high $x$. Thus, there is some tension between the the Run I and the Run II Tevatron jet data. This tension was examined by CTEQ~\cite{Pumplin:2009nk}; although some degree of tension does exist, the data sets from Run I and Run II were found to be statistically compatible with each other, with the tensions similar to that between other data sets used in the global fit. Thus, both generations (Run I and Run II) of Tevatron jet data have been kept in recent CTEQ PDF fits~\cite{CTEQ10}, although only the Run II data have been used in the MSTW and NNPDF global fits. The resultant high $x$ gluon distribution for CTEQ is thus larger than either MSTW or NNPDF. 
MSTW has also found the impact 
of the Run II jet data on the uncertainty in the large-$x$ gluon distribution 
obtained in the global PDF fits to be relatively minor (see Figure~\ref{fig:gluon}), while CTEQ finds the constraints to still be appreciable. 

A substantial fraction of jets produced in the forward direction results from 
the simultaneous scattering of two pairs of partons. The rate for such processes is defined by 
double-parton distribution functions (dPDFs), i.e. 
the probability to find two partons with  certain momenta inside the nucleon. 
Due to experimental constraints that are currently insufficient, 
the dPDFs are usually derived as a product of 
conventional collinear PDFs. However, the dPDFs obtained in such a way 
do not enjoy the fermion and momentum sum rules fulfilled for the 
dPDFs evolution equations. Addressing this shortcoming,
a novel set of dPDFs 
was generated with the MSTW08 PDFs taken as an input~\cite{Gaunt:2009re}.

For many important processes the fixed-order perturbative QCD 
calculations demonstrate excellent convergence~\cite{Gehrmann:2010rj} and 
in most cases the NNLO approximation is sufficient 
to meet the data accuracy. However, at small $x$ the perturbative corrections
are unstable. In particular,  
the 3-loop terms in the massless DIS coefficient functions are 
quite large~\cite{Vermaseren:2005qc}. On the other hand,
the small-$x$ resummation of the splitting functions and the DIS 
Wilson coefficients softens the small-$x$ terms and  
compensate to some extent the NNLO 
corrections~\cite{Ciafaloni:2007gf, Altarelli:2008aj}.
The impact of small-$x$ resummation in DIS
was examined  with a variant 
of the NNPDF1.0 PDFs~\cite{NNPDF1} fitted to the high-$Q$ part of the 
DIS data, and then extrapolated to the low-$Q$ region
with the heavy-quark contribution calculated  
in the ZMVFN scheme~\cite{Caola:2009iy}.
In this way, possible deviations from the standard NLO evolution of the 
inclusive HERA data were found. 
The effect observed is explained
in Ref.~\cite{Caola:2009iy} by the impact of  small-$x$ resummation 
or by parton saturation~\cite{Gribov:1984tu}. 
In line with this observation, 
the GMVFN variant of the HERAPDF fit of Ref.~\cite{radescu1}
is sensitive to the low-$Q$ part of the HERA data; however the 
FFN variant of the fit is much more stable to the cut on $Q$. 
Resummation effects in DIS thus still need to be systematically 
explored. This is an important issue for collider phenomenology 
since  resummation contributes to many important collider channels such as 
heavy-quark and the lepton pair production (cf. 
Ref.~\cite{Diana:2010ef}  for a recent study of the effects of resummation 
in direct photon production). Small-$x$ dynamics
is often considered within the framework of 
non-collinear PDF evolution, using  $k_T$ or angular ordering. 
The unintegrated parton distributions  appear in the parton evolution 
with the angular ordering depending on both longitudinal and 
transverse variables; therefore, the calculation of 
final state transverse momentum distributions is better suited for this formalism. With the appearance 
of the combined HERA I data, a determination of the unintegrated gluon distribution was updated
and found enhanced as compared to the previous 
determination~\cite{knut}. 

PDF shapes cannot be calculated in perturbative QCD. Instead, usually 
they are parameterized in a model-independent way 
with  loose constraints imposed on the high-$x$ and the low-$x$ 
PDF exponents, coming from quark counting rules and  Regge 
phenomenology, respectively~\cite{MSTW,CTEQ,JR,ABKM,HERACOMB}. 
If the PDFs are evolved starting from a scale $\mu_{\rm F}^0<1~{\rm GeV}$, 
like in the case of JR PDFs, at the hard process scales they 
enjoy the asymptotic behavior defined by the leading evolution 
kernel singularities~\cite{derujula}. 
In a similar way, the non-singlet PDF combinations can be  
constrained by the infrared QCD evolution kernel~\cite{ermolaev}.
An additional constraint may come from unitarity~\cite{tiwon}
and the universalities observed for the  
proton, photon, and diffractive structure functions~\cite{klim}.
\begin{wrapfigure}{hr}{0.5\columnwidth}
\includegraphics[width=20pc,height=13pc]{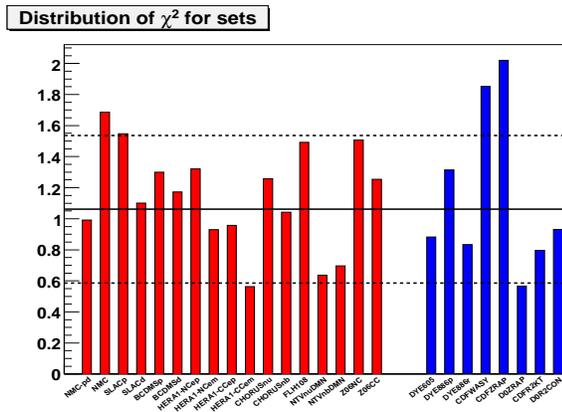}
\caption{Distribution of the $\chi^2/NDP$ over the data sets used in the 
NNPDF2.0 fit. Figure is taken from Ref.~\cite{NNPDF}.
}
\label{fig:chi2}
\end{wrapfigure}
In contrast, the NNPDF collaboration uses a very flexible form of PDF parameterisation,
with 259 free parameters~\cite{NNPDF}.
The gradient minimization methods are inapplicable for 
that large a  number of parameters, and instead the 
NNPDF PDFs were fitted to the 
data using the neural network technique.  
Within the neural network approach, the PDF uncertainties are 
calculated from the probability distribution of the neural network 
replicas. Since no tensions between the different 
data sets used in the fit were observed (see Figure~\ref{fig:chi2}) the 
standard statistical methods are employed for the NNPDF PDFs.
The uncertainties in the HERAPDF, ABKM, and JR PDFs are also 
calculated using the standard statistical methods. 
For comparison, the CTEQ and MSTW collaborations apply larger  
tolerance factors to take into account discrepancies 
between the data sets used in the fit and theoretical uncertainties, such as parameterisation choices. 
Despite the different statistical treatments, 
the PDF errors provided by all groups are in  qualitative agreement with each other.

Outside the  kinematic region covered by the existing data,
the NNPDF PDF uncertainties are larger than the other PDF uncertainties,  
due to the lack of prior theoretical/parameterisation constraints on the PDFs. 
On the other hand,  the uncertainties for the JR PDFs
are reduced, compared to other fits 
due to additional theoretical constraints.
A study of the small-$x$
sea and gluon distribution flexibility allowed by the data 
was also performed with the PDFs parameterized by the Chebyshev 
polynomials~\cite{radescu2,Pumplin:2009bb}. The gluon distribution 
obtained in Ref.~\cite{radescu2} is stable with respect to the 
polynomial powers used at $x\gtrsim 0.001$; at smaller $x$ the gluon 
distribution is unstable, due to the lack of the experimental constraints.
%

The PDFs given in Table~\ref{tab:pdfs} are similar, but do not completely overlap within $1\sigma$ uncertainty bands.
This discrepancy appears to be due to differences in the theoretical formalisms used in the global fits 
and  in details of the data treatment;  however, no definitive understanding has yet been reached. 
In addition, the treatment of the strong coupling 
constant $\alpha_{\rm s}$ is different for the PDFs listed in  Table~\ref{tab:pdfs}.

\begin{wrapfigure}{hr}{0.5\columnwidth}
\hspace{-0.3cm}
\includegraphics[width=19pc,height=14pc]{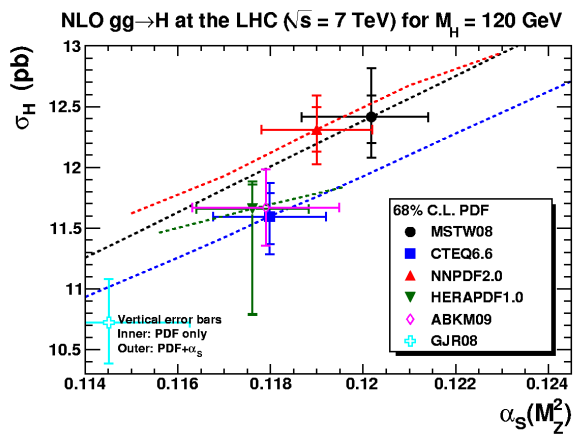}
\caption{Comparison of the NLO Higgs rates at the LHC 
collision energy of 7 TeV calculated with 
different PDFs. Figure is taken from Ref.~\cite{watt}.
}
\label{fig:higgs}
\end{wrapfigure}
In Refs.~\cite{MSTW,JR,ABKM} the value of $\alpha_{\rm s}$ 
is fitted to the data 
simultaneously with the PDFs, while in the fits of 
Refs.\cite{CTEQ,NNPDF,HERACOMB} it is fixed at a value close to the 
world average~\cite{Bethke:2009jm}. For many cross sections, such
as Higgs and top quark pair production, 
the rates are quite sensitive to the value of $\alpha_{\rm s}$ used
(cf. Refs.~\cite{Martin:2009bu,Lai:2010nw,Demartin:2010er}). 
This sensitivity, and the variations in the values of $\alpha_s$ used, results in  additional uncertainties on the hard-scattering cross section
predictions. For the NLO Higgs cross section benchmark,
the spread of the predictions is as large as 20\% (see Figure~\ref{fig:higgs}).  
Further
consolidation of the predictions would make LHC predictions for the 
Higgs cross section more definitive. The  corresponding spread for predictions at the Tevatron can also be important for the interpretation of the 
Fermilab collider data~\cite{Baglio:2010um}. 

The PDF4LHC working group is carrying out a benchmarking exercise~\cite{huston} where each PDF group has been invited to provide NLO predictions for benchmark processes (such as the
Higgs production cross sections for the Higgs masses of 120, 180 and 240 GeV, 
as well as for $W,Z$ and $t\bar{t}$ production. The predictions are to be made for the default 
value of $\alpha_s$, as well as for a range of values from 0.116 to 0.118. 
Comparisons are available at the website~\cite{pdf4lhc},
along with a prescription for the calculation of the PDF uncertainties at the LHC.
\\ \\

The last year has seen tremendous progress on both the theoretical and the experimental fronts. In particular, the release of the combined H1+ZEUS HERA I data has led to increased precision and thus better constraints on PDFs, especially the low-$x$ gluon distribution. The Tevatron Run II W-lepton asymmetry measurements have the potential to allow for improvements in the description of 
high-$x$ quark distributions, but conflict with some of the data sets currently in use in the global fits. A resolution of the conflict, perhaps with better understanding of some of the nuclear corrections for the fixed target data that conflict with the lepton asymmetry data, is needed.

The LHC is presenting its first results for standard model cross sections. Predictions for these cross sections are available using PDFs from a number of fitting groups. It is useful and important at this stage of LHC running to have benchmark comparisons of the predictions from the various PDF groups. Such an exercise has been carried out by the PDF4LHC working group.
Further standardization would be extremely useful, 
in particular common estimate of the uncertainty in 
the value of $\alpha_s$.

At the next DIS workshop, we expect our first inputs from LHC data to PDF fitting. To significantly affect the existing PDF fits, the systematic errors must be reasonably small and well-known for the measurements to be included. The most likely cross sections to be used in such a way are Drell-Yan measurements, especially of $W$ and $Z$ boson production, with event yields of the order of a million events per experiment to be expected for 1 $fb^{-1}$, and with the possibility of accessing new kinematic regions in $x$ and $Q^2$\cite{keaveney,lorenzi,dahmes}. 
\\

%





\end{document}